# Computational model of drug dissolution in the stomach: effects of posture and gastroparesis on drug bioavailability

Jae H. Lee[1,2†], Sharun Kuhar[1], Jung-Hee Seo[1,2], Pankaj J. Pasricha[3], and Rajat Mittal[1,2,4‡]

[1]Department of Mechanical Engineering, Johns Hopkins University, Baltimore, MD, USA

[2]Institute for Computational Medicine, Johns Hopkins University, Baltimore, MD, USA

[3]Division of Gastroenterology and Hepatology, Johns Hopkins School of Medicine, Baltimore, MD, USA

[4]Department of Medicine, Johns Hopkins School of Medicine, Baltimore, MD, USA

## Abstract

The oral route is the most common choice for drug administration because of convenience, low cost, and high patient compliance, but is also a complex route. The rate of dissolution and gastric emptying of the dissolved active pharmaceutical ingredient (API) into the duodenum is modulated by factors such as gastric motility, but current in-vitro procedures for assessing drug dissolution are limited in their ability to recapitulate this process. This is particularly relevant for disease conditions, such as gastroparesis, that alter the anatomy and/or physiology of the stomach. In this study we employ a biomimetic in-silico simulator based on the realistic anatomy and morphology of the stomach, to investigate the effect of body posture and stomach motility on drug bioavailability. The simulations show that changes in posture can have a significant (up to 83%) effect on the emptying rate of the API into the duodenum. Similarly, reduction in antral contractility associated with gastroparesis can also significantly reduce the dissolution of the pill as well as emptying of the API into the duodenum. The simulations show that for an equivalent motility index, reduction in gastric emptying due to neuropathic gastroparesis is larger by a factor of about five compared to myopathic gastroparesis.

**Keywords:** computational fluid dynamics, immersed boundary method, fluid-structure interaction, gastric motility, oral drugs

---

† **Present address:** Center for Drug Evaluation and Research, U.S. Food and Drug Administration, Silver Spring, MD, USA.

‡ **Corresponding Author:** mittal@jhu.edu

## 1. Introduction

The oral administration route is a safe, economic, and easy way to administer drugs to patients and one that is known to result in a high degree of patient compliance[1]. However, the oral route is actually the most complex way for an active pharmaceutical ingredient (API) to enter and be absorbed by the body. The bioavailability of the drug in the gastrointestinal (GI) tract depends not only on the drug formulation, but also on the dynamic physiological environment in the fed stomach[21]. This environment arises from the complex interplay of factors such as the contents of the stomach, stomach motility, and the gastric fluid dynamics. In particular, stomach contractions induce pressure and shear forces that generate complex pill trajectories. This results in varying rates of pill dissolution and non-uniform emptying of the drug into the duodenum. This sometimes lead to phenomenon such as "gastric dumping" in the case of modified-release dosage forms[21]. These issues pose several challenges to the design of drugs delivery systems in R&D, clinical, and regulatory settings.

One of the important properties to evaluate for oral drugs is the rate of dissolution, which is often the rate limiting step in drug absorption[29]. Existing approaches to assessing/quantifying drug dissolution rely primarily on in-vitro models but recapitulating the conditions experienced by an oral drug formulation in the stomach via these in-vitro models has significant limitations. The US Pharmacopeia (USP) dissolution apparatus (I-IV) are the de-facto standard (particularly, USP II) for assessing and quantifying drug dissolution (Fig. 1A), but a variety of studies have shown the significant shortcomings of these devices for mimicking the conditions of the stomach[11]. More advanced in-vitro models[20,25] have attempted to mitigate these shortcomings but at the cost of increasing device complexity. Furthermore, despite the increased complexity, these in-vitro simulators are still unable to adequately recreate bio-relevant conditions of motility-induced mixing, shear and pressure, the biochemical status associated with food content and gastric secretions[4,11].

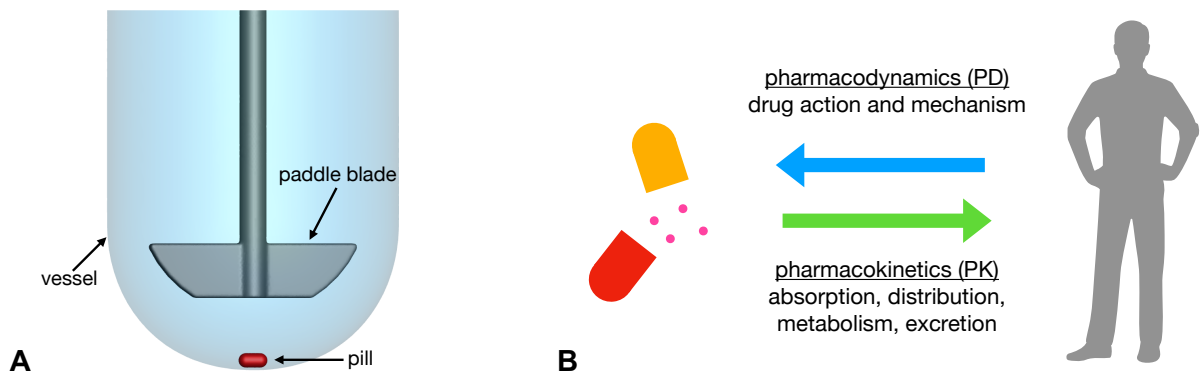

**Fig. 1**. **A:** A schematic of the US Pharmacopeia apparatus type II (USP II: paddle). **B:** A schematic of pharmacokinetic/pharmacodynamic (PK/PD) model for dose-effect analysis and assessing physiological interaction and metabolism of drugs in vivo[26].

Another widely used approach to quantifying drug release and absorption in the early phase of drug development is pharmacokinetic/pharmacodynamic (PK/PD) analysis (Fig. 1B), which is a technique used to quantify and study the effect of physiological interaction and metabolism of drugs in vivo[26]. PK/PD analysis is robust in that it not only provides "dose-effect" analysis of the drug, but it also gives information on how much of the administered dose is delivered to specific receptors, how much is metabolized, and the variability in drug effect between subjects[26]. Despite its advantages and development of more complex PK/PD models, PK/PD analysis is still unable to capture some of the key factors that affect bioavailability of oral drugs, such as posture and gastric motility[29].

Several studies have shown that differences in posture yield considerable level of variations within and between subjects in gastric mixing and emptying, digestion, and absorption of liquid and solid meals, as well as in drug bioavailability[17,27]. The effect of posture on drug bioavailability is assessed pharmacokinetically by measuring, for example, time to reach peak plasma drug concentration, maximum plasma drug concentration, and total exposure. However, the exact mechanism and the magnitude of postural effects have not been understood completely[27].

These challenges are particularly relevant for many disease conditions that are associated with alterations in the anatomy and/or physiology of the stomach. For example, gastroparesis is a chronic change in  gastric motility caused by conditions such as diabetes, Parkinson's disease, collagen vascular disorders, and Roux-en-Y gastric bypass[24]. Clinical studies such as PK/PD, however, involve healthy volunteers with an assumption that the inter-individual differences in GI physiology is small[14]. Current preclinical and clinical approaches to assess the efficacy of oral drugs are limited in elucidating the relationship between various diseased states that alter gastric motility and also in accounting for other relevant parameters such as the volume, composition, and fluid dynamics of the gastric content[29]. Therefore, the quantitative data that incorporates these variations in gastric functions, especially in diseased states, are lacking.

Computational fluid dynamics (CFD) is a powerful complement to benchtop experiments that has been widely used in developing and and/or assessing medical devices such as ventricular assist devices[30], prosthetic heart valves[23], and blood filters[8]. Indeed, in-silico modeling and simulation have become a strategic priority for the U.S. Food and Drug Administration in the recent years in supporting regulatory evaluation of biomedical products[23]. CFD can enable us to obtain full flow information in anatomically realistic stomach to investigate the effect of the interaction between gastric biomechanics and fluid dynamics in drug dissolution and determining the local drug concentrations. Using an in-silico simulator, we can easily test different physiological and other conditions (such as posture) in parallel, and also generate databases that more closely and broadly represent the physiology of different gastric conditions. In this study, we leverage a computational modeling platform to simulate drug dissolution in an image-based human stomach model. The models are used to investigate the effect of posture and gastroparesis on drug dissolution in the stomach as well as the release of the drug into the duodenum.

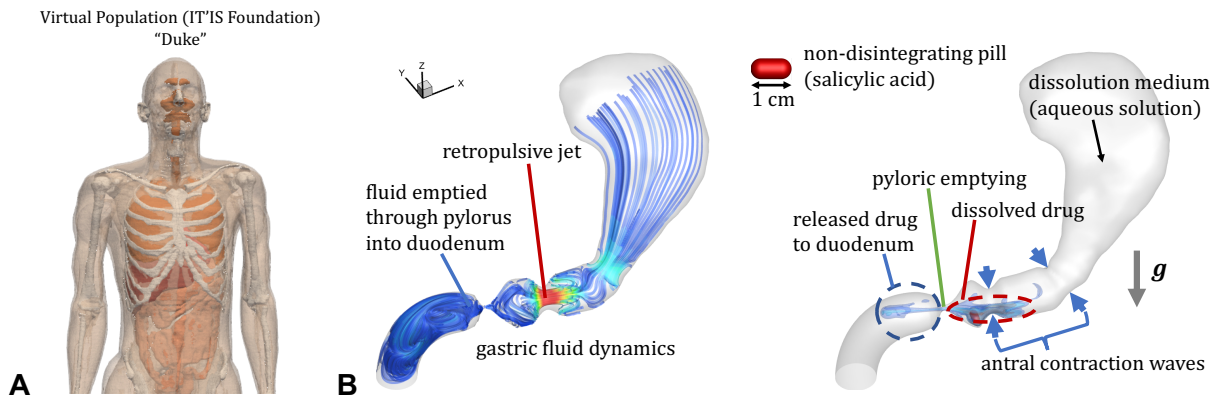

**Fig. 2. A.** The anatomy of the human stomach is obtained from the Virtual Population library[13]. For this study, we use the stomach from "Duke", a 34-year-old male adult. **B.** Three-dimensional stomach model is then created by segmenting the stomach lumen.

## 2. Materials and Methods

### 2.1 Stomach model

An anatomical model of human stomach is constructed from in-vivo imaging data available from the Virtual Population library[13]. The database provides high resolution anatomical models created from magnetic resonance image data. For this study, we use the stomach model from "Duke" (Fig. 2A), who is a 34-year-old male adult. The 3D stomach lumen is segmented and reconstructed from the MRI dataset as unstructured surface mesh with triangular elements.

In our stomach model, we focus on the dissolution, mixing, and emptying driven by antral contractions associated with a fed stomach, and this current model does not include tonic contractions. Antral contractions are modeled as a series of pulse waves called antral contraction waves (ACWs) propagating from the corpus region and all the way down to the terminal antrum. The terminal antral contraction (TAC) is a segmental contraction of the terminal antrum, which is activated when the ACW reaches the terminal antrum[6]. The pylorus is modeled in a similar fashion with a prescribed motion, in which the pyloric closure is controlled by the start-time and duration of the closure. The pyloric sphincter is closed for 65% of the time in one cycle of antral contraction, and the pyloric closure and the TAC begin synchronously as the ACW reaches the distal antrum region, similar to the healthy model described by Ishida et al.[16] The diameter of the pylorus is approximately 2 mm when open during fed state[19].

The pill is assumed to be made of salicylic acid, with a specific gravity of $\rho_p/\rho = 1.2$ [2] and modeled as a non-disintegrating and non-deformable object, an approximation that is valid for the early stages in the dissolution of the pill. Because the pill is denser than the fluid medium, the pill settles down on the stomach wall due to gravity. Here, we assume that the fluid medium in the stomach is homogeneous (e.g. water, juice, and milk). As the drug dissolves, the dissolved API is transported by the gastric flow that is driven by the antral contractions. The coordination between the antral contraction and the pylorus opening and closing results in a pulsatile release of the stomach contents, including the API into the duodenum. Fig. 2B shows the schematic diagram of the features in our model and further details of the mathematical formulation and governing equations of our model can be found in the Supplementary Materials.

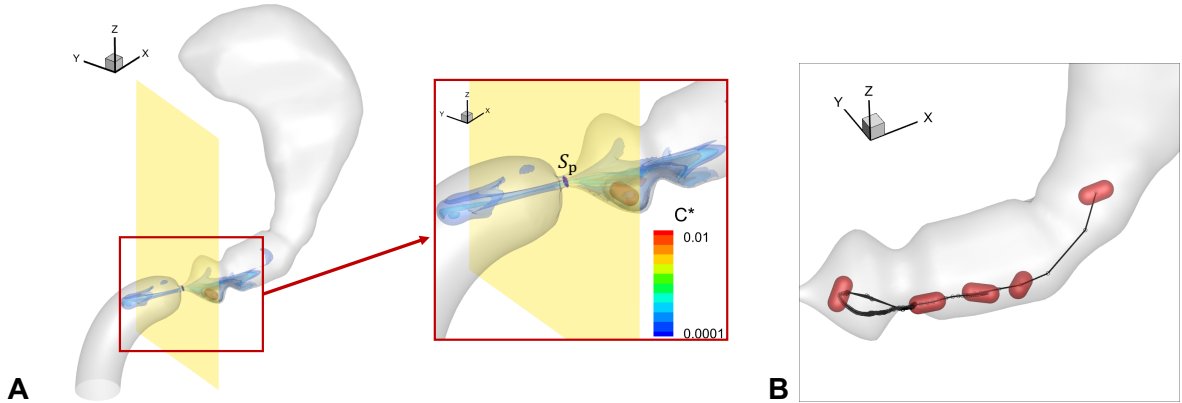

**Fig. 3. A.** Computational model of the human stomach and duodenum system in this study. The cross-section of the pylorus (purple circular plane in the pylorus denoted as $S_p$ formed with the plane shown is where the rate of gastric emptying and the dissolved API concentration into the duodenum through the pylorus is measured. **B**. Our computational model includes free motion of the pill induced by the motility of the stomach, the gastric fluid dynamics, and gravity. The trajectory of the pill can be traced (black line).

This in-silico platform enables us to observe and quantify pill dissolution in the stomach and as well as the release of the API into the duodenum. We estimate the emptying rate of the overall liquid content in the stomach, $Q_{\text{emptying}}$, as well as the dissolved mass into the duodenum by calculating the mass transfer rate through the pylorus, $\dot{m}_{\text{API}}$,

$$Q_{\text{emptying}} = \int_{S_p} \boldsymbol{u} \cdot \boldsymbol{n} \ \mathrm{d}A, \qquad \dot{m}_{\text{API}} = \int_{S_p} C_{\text{API}} \boldsymbol{u} \cdot \boldsymbol{n} \ \mathrm{d}A, \qquad (1)$$

in which $C_{\text{API}}$ is the concentration of the dissolved API, $S_p$ is the cross-section of the pyloric opening, and $\boldsymbol{n}$ is a unit normal to the plane shown in Fig. 3A. Our model also allows us to trace the exact trajectory of the pill motion (Fig. 3B), which is induced by a complex interplay of stomach motility, gastric fluid dynamics, and gravity.

Another quantity that we can look at is the total amount of API released into the duodenum every cycle of antral contraction. We can use this data to fit a modified Elashoff's model[19], a sigmoid-shape function that has been used to describe gastric emptying rate. This model can provide insight into the long-term behavior for different conditions. In this model, the amount of API being emptied per cycle, $\mathrm{API}(t)$, is described as

$$\mathrm{API}(t) \;=\; \mathrm{API}_\infty (1 - e^{-\alpha t})^\beta, \tag{2}$$

in which $\mathrm{API}_\infty$ is the predicted amount of API released per cycle at steady-state (in mg), and $\alpha$ and $\beta$ represent emptying rate and initial delay in emptying, respectively.

## 3. Results

### 3.1 Effect of body posture on drug dissolution and release

An advantage of using an in-silico simulator is that we can tightly control the operating conditions, eliminating variations other than the variable of interest between different cases. To compare the effect of posture on the bioavailability of the drug, we consider four different postures: upright, leaning right, leaning left, and leaning back with the rest of the parameters kept constant. To model postural effects, we keep unchanged the stomach geometry, the initial position of the pill, and the gastric biomechanics, and instead of actually rotating the stomach as shown in Fig. 4, we just change the direction of the gravitational force as described in Table 1. Because the pill is denser than the dissolution medium, the effect of the direction will significantly affect the pill motion, and thereby the rate of dissolution and release of the API. Fig. 5 shows the time-dependent volumetric distributions of the dissolved API concentrations in the antral and duodenal regions for different postures.

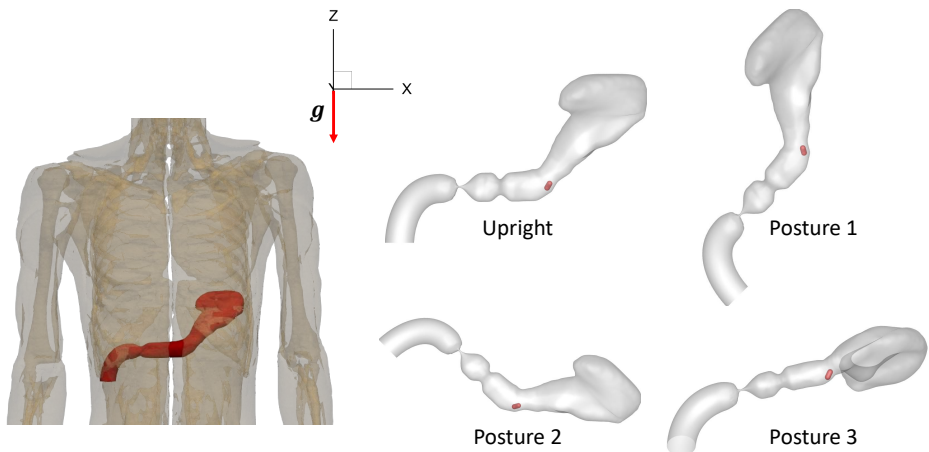

**Fig. 4.** Diagram showing the original position of the stomach relative to the body and different relative positions of the stomach with respect to the direction of gravity considered in this study.

|  | upright | posture 1 (leaning right) | posture 2 (leaning left) | posture 3 (leaning back) |
|---|---|---|---|---|
| direction of gravity | $(0,0,-1)$ | $(-\sqrt{2}/2, 0, -\sqrt{2}/2)$ | $(\sqrt{2}/2, 0, -\sqrt{2}/2)$ | $(0, \sqrt{2}/2, -\sqrt{2}/2)$ |

**Table 1.** Different postures considered in this study. Here, the stomach geometry is fixed in space, and the direction of gravity is changed. The unit vectors indicate different directions of gravity considered.

The upright position (Fig. 5A) shows a typical behavior in which the pill is carried by the ACWs toward the pylorus and moved away from the pylorus by the combination of the retropulsive jet[7] and the antral relaxation[16]. When the body is leaning right by 45° (Fig. 5B), we see a significant increase in the dissolved API in the stomach and the duodenum during this early time. We know that the emptying of the dissolved API into the duodenum is proportional to the API concentration at the pylorus (Eq. 1), and the concentration at the pylorus is high in this case. The reason for this increased concentration at the pylorus

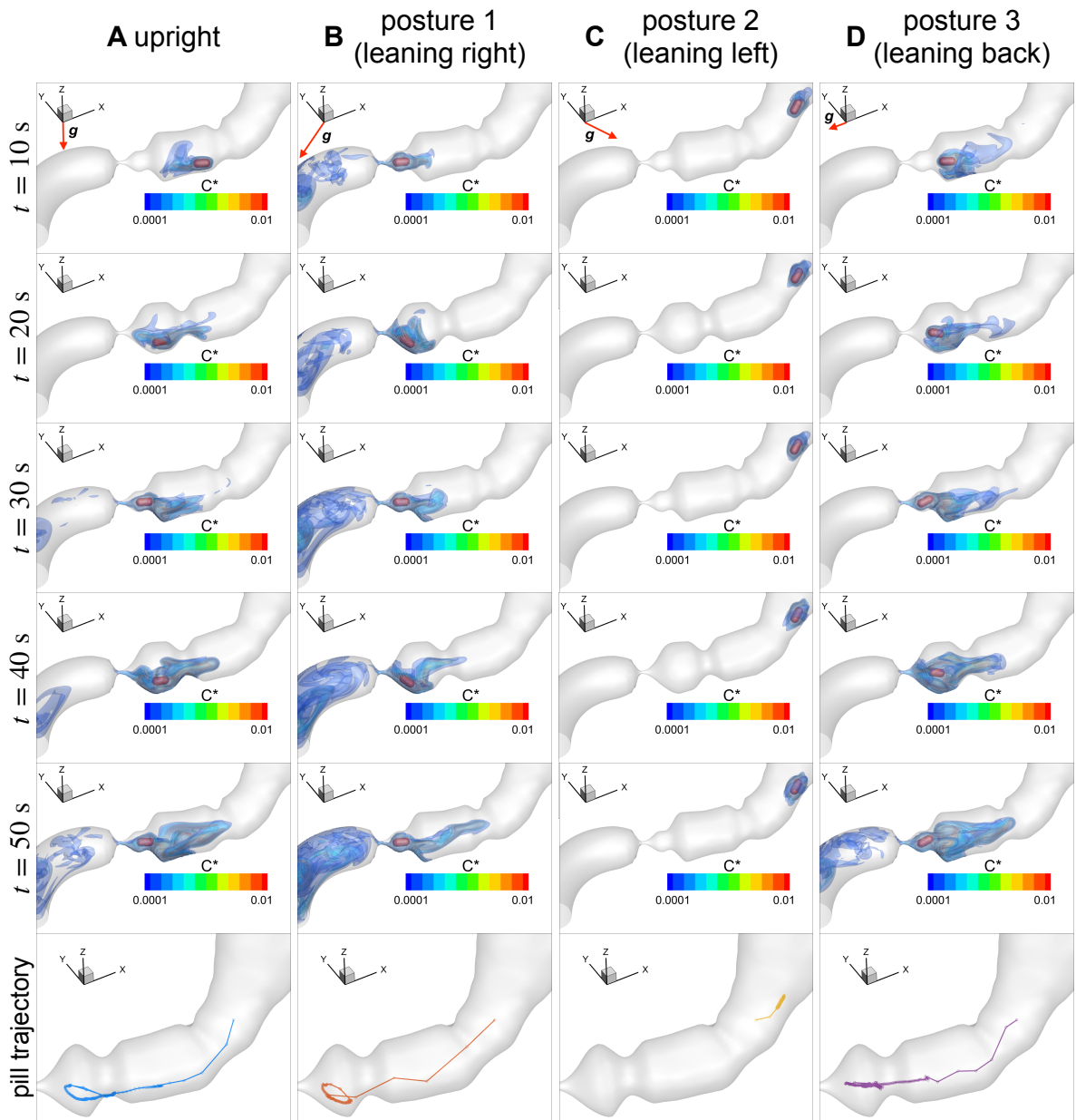

**Fig. 5.** Volumetric distributions of the dissolved active pharmaceutical ingredient (API) concentrations in the antrum and duodenum regions for different postures ($C^* = C_{API}/C_s$ is the normalized concentration, and $C_s$ is the solubility concentration of the pill). Red arrows show the relative direction of gravity. The last row shows the pill trajectory for each case.

is that the pill resides close to the pylorus throughout the duration of the simulation because the gravity is now pulling the pill along the antrum toward the pylorus. On the other hand, when leaning left by 45° (Fig. 5C) the pill is kept in a position away from the antrum and the large contractions from the traveling ACWs. This results in the dissolved

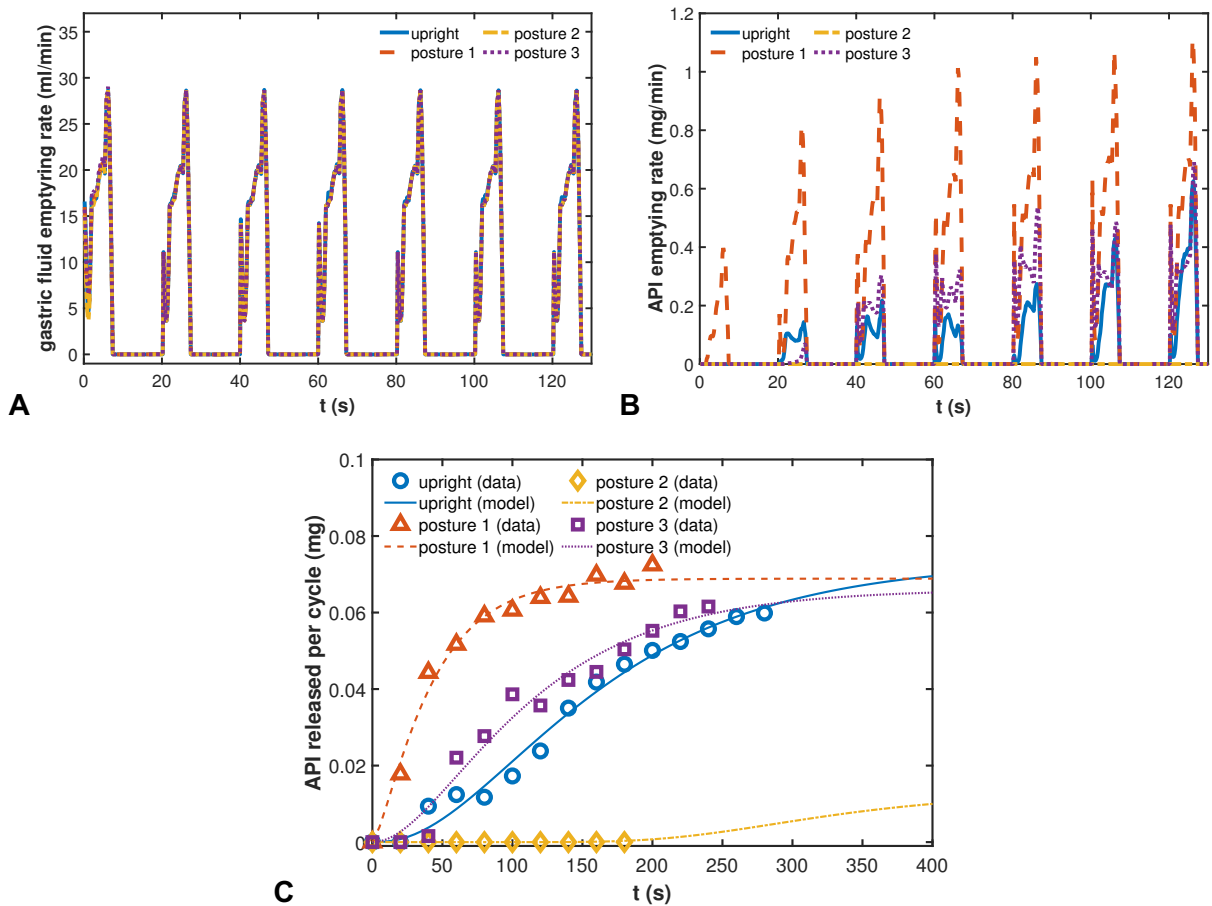

**Fig. 6.** Comparison for different postures. **A.** Quantification of gastric emptying rate into the duodenum measured by calculating the velocity flux through the pylorus. **B.** Quantification of the dissolved mass into the duodenum measured by calculating the concentration flux through the pylorus. **C.** Comparison of fit for a modified Elashoff model against simulation data for the amount of API emptied into the duodenum every cycle for different postures.

API concentration at the pylorus and in the duodenum being close to zero. Lastly, when the body is leaning back by 45° (Fig. 5D), a larger amount of API is released because the pill is closer to the core of the jet and experiences stronger retropulsive jet. The retropulsive jet is known to be responsible for rapid gastric mixing[19], which accelerates the release of the API from the pill. The direction of gravity also pulls the pill slightly more towards the distal antrum region more than the upright position does. This eventually results in larger amount of API concentration dissolved in the pylorus region and thereby increase in the dissolved API released into the duodenum (Fig. 6B).

| | upright | posture 1 (leaning right) | posture 2 (leaning left) | posture 3 (leaning back) |
|---|---|---|---|---|
| average dissolved API released | 0.016 mg | 0.052 mg | 0.0 mg | 0.024 mg |

**Table 2.** Average dissolved active pharmaceutical ingredient (API) released into the duodenum per cycle for different postures. To calculate the average dissolved API released, the cycle-average of the flux of dissolved API in Fig. 6B is calculated, which is then integrated over the duration of one cycle (20 sec).

Fig. 6A indicates that gravity does not affect gastric emptying of the liquid content of the stomach, which confirms the clinical findings of Golub et al. that the rate of emptying of the liquid only depends on the volume[12]. Fig. 6B shows that leaning right (posture 1) yields significant increase in the dissolved mass released into the duodenum compared to other positions. Leaning left (posture 2) leads to significant decrease.

Table 2 summarizes our findings in terms of average dissolved API released into the duodenum per cycle for each posture. To calculate this, we first obtain the cycle-average of the flux of dissolved API shown in Fig. 6B, which is then integrated over the duration of one cycle (20 sec). We note that for the upright position, 0.016 mg of API is released into the duodenum per cycle, 0.052 mg for posture 1 (leaning right), 0.0 mg for posture 2 (leaning left), and 0.024 mg for posture 3 (leaning back). Here, the mean dissolved API is 0.023 mg and the root mean squared deviation (RMSD) among the four simulated cases is 0.019 mg.

Fig. 6C compares the simulation data for the amount of API emptied into the duodenum every cycle for different types of gastroparesis. The data is then used to fit the modified Elashoff model (Eq. 2). The fitted parameters $\mathrm{API}_\infty$, $\alpha$, and $\beta$ are listed in Table 3. The predictions from the model shows that the difference in the amount of API being emptied per contraction between the upright, leaning right, and leaning right positions are eventually similar in long-term ($\mathrm{API}_\infty$). However, it is clear that depending on the posture there are significant differences in the rate of emptying and initial delay in emptying.

| | upright | posture 1 (leaning right) | posture 2 (leaning left) | posture 3 (leaning back) |
|---|---|---|---|---|
| $\mathrm{API}_\infty$ (mg) | 0.0735 | 0.0689 | 0.0131 | 0.0661 |
| $\alpha$ | 0.0096 | 0.0284 | 0.0116 | 0.0126 |
| $\beta$ | 2.6004 | 1.4225 | 27.7228 | 2.1235 |

**Table 3.** Parameters obtained by fitting the modified Elashoff model (Eq. (2)) against the simulation data for the amount of API emptied into the duodenum every cycle for different postures (Fig. 6C). $\mathrm{API}_\infty$ is the predicted amount of API released per cycle at steady-state, and $\alpha$ and $\beta$ represent emptying rate and initial delay in emptying, respectively.

### 3.2 Effect of gastroparesis on drug dissolution and release

Gastroparesis is typically identified with neuropathic and/or myopathic abnormalities that decrease the number and/or the amplitude of contractions, respectively[22]. Antral motility is often assessed quantitatively via a motility index (MI), which is defined as

$$\mathrm{MI} = [\text{mean amplitude}] \times [\text{frequency of contractions every 3 minutes}],$$

and gastroparesis is identified with a decrease in MI[10,22]. Gastroparesis can be then further categorized into neuropathy (damaged vagus nerve), myopathy (increased fibrosis or degeneration of smooth muscle cells), or both[5]. In our simulations, we characterize different types of gastroparesis as listed in Table 4, in which the frequency of ACW is reduced to model a case of neuropathy ($T_\mathrm{p}$ = 40 sec; MI = 2.0) and the amplitude of ACW is reduced to model myopathy ($\lambda_\mathrm{a,max}$ = 0.2; MI = 1.8), and both ($T_\mathrm{p}$ =

| | normal | neuropathy | myopathy | neuropathy+ myopathy |
|---|---|---|---|---|
| $\lambda_{a,max}$ | 0.45 | 0.45 | 0.2 | 0.2 |
| $T_p$ | 20 sec | 40 sec | 20 sec | 40 sec |

**Table 4.** Different gastric motility parameters that model different types of gastroparesis considered in this study. Gastroparesis is identified with neuropathic and/or myopathic abnormalities that decrease the number ($\propto 1/T_p$) and/or the amplitude of contractions ($\lambda_{a,max}$), respectively.

40 sec and $\lambda_{a,max} = 0.2$; MI = 0.9). In the cases we consider, we fix the posture of each case to be upright, so the healthy normal case is the same as the upright case in Section 3.1.

Fig. 7 shows time-dependent volumetric distributions of the dissolved API concentrations in the antral and duodenal regions for different types of gastroparesis. The neuropathic case (Fig. 7B) shows that the dissolved API concentration is distributed further up away from the pylorus although the strength of the retropulsive jet is unchanged. This is because of a delay in the arrival of the next ACW to push the API concentration back toward the pylorus against the increased duration of retropulsive jet. Although the dissolved mass is still well-mixed despite the delayed arrival of successive ACWs with the help of sustained strong antral contraction, this delay results in reduced transport of the dissolved API into the duodenum. In the case of gastric myopathy (Fig. 7C), the amplitude of the antral contraction is reduced to less than half that of the normal case. The dissolved API is pushed toward the pylorus as frequently as that in the normal case. However, the strength of the retropulsive jet is decreased and most of the API concentration therefore resides closer to the pylorus compared to the normal and the neuropathic cases. Despite this increased API concentration residing closer to the pylorus, the API is not transported to the pylorus and released into the duodenum as efficiently as it did in the normal case. In the stomach with combined neuropathy and myopathy (Fig. 7D), the effect of gastric motility is significantly reduced. Therefore, the dissolution and transport of the API is driven primarily via diffusion, with some mixing induced by the retropulsive jet.

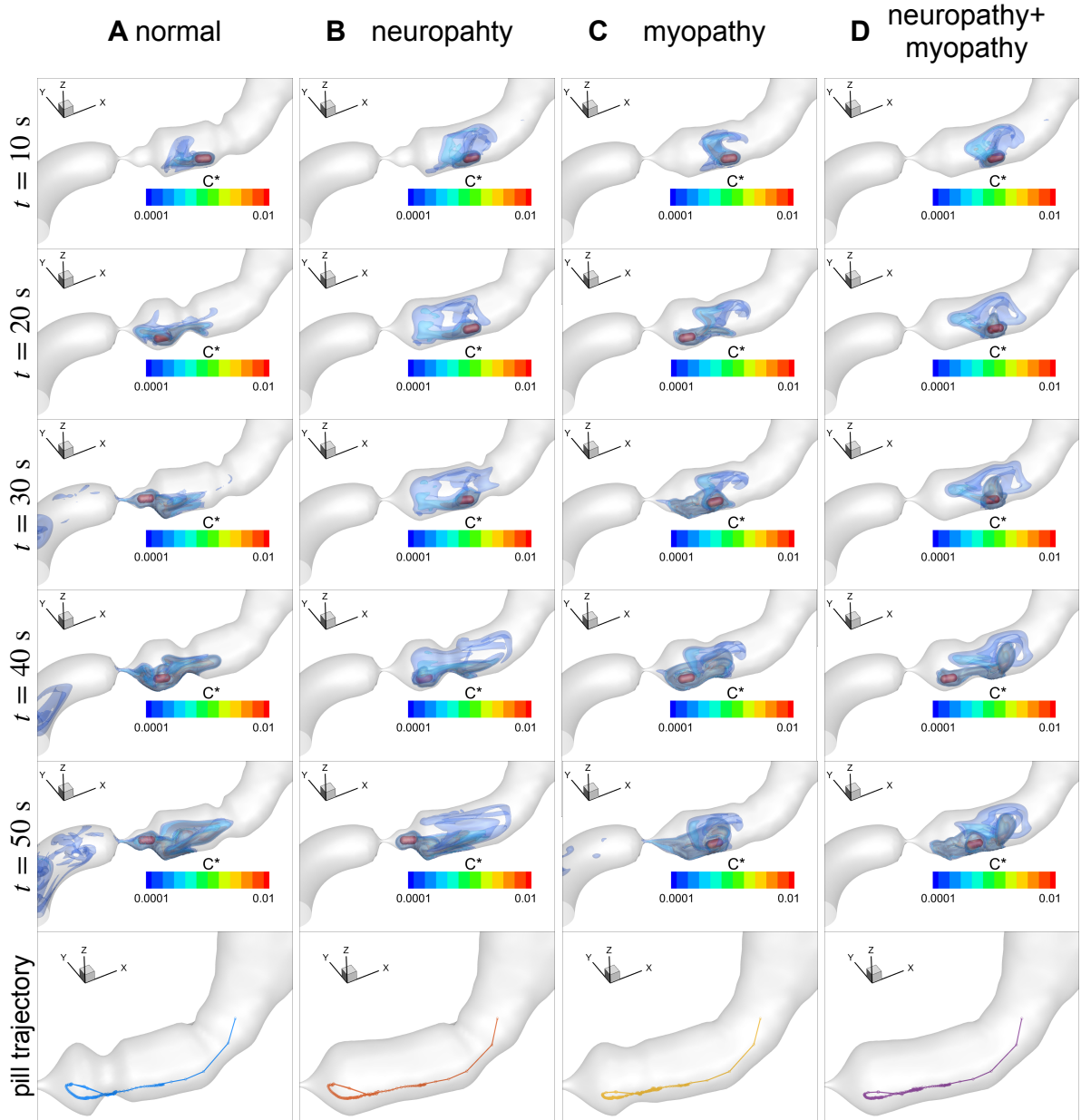

**Fig. 7.** Volumetric distributions of the dissolved active pharmaceutical ingredient (API) concentrations in the antrum and duodenum regions for different types of gastroparesis ($C^* = C_{API}/C_s$ is the normalized concentration, and $C_s$ is the solubility concentration of the pill). The last row shows the pill trajectory for each case.

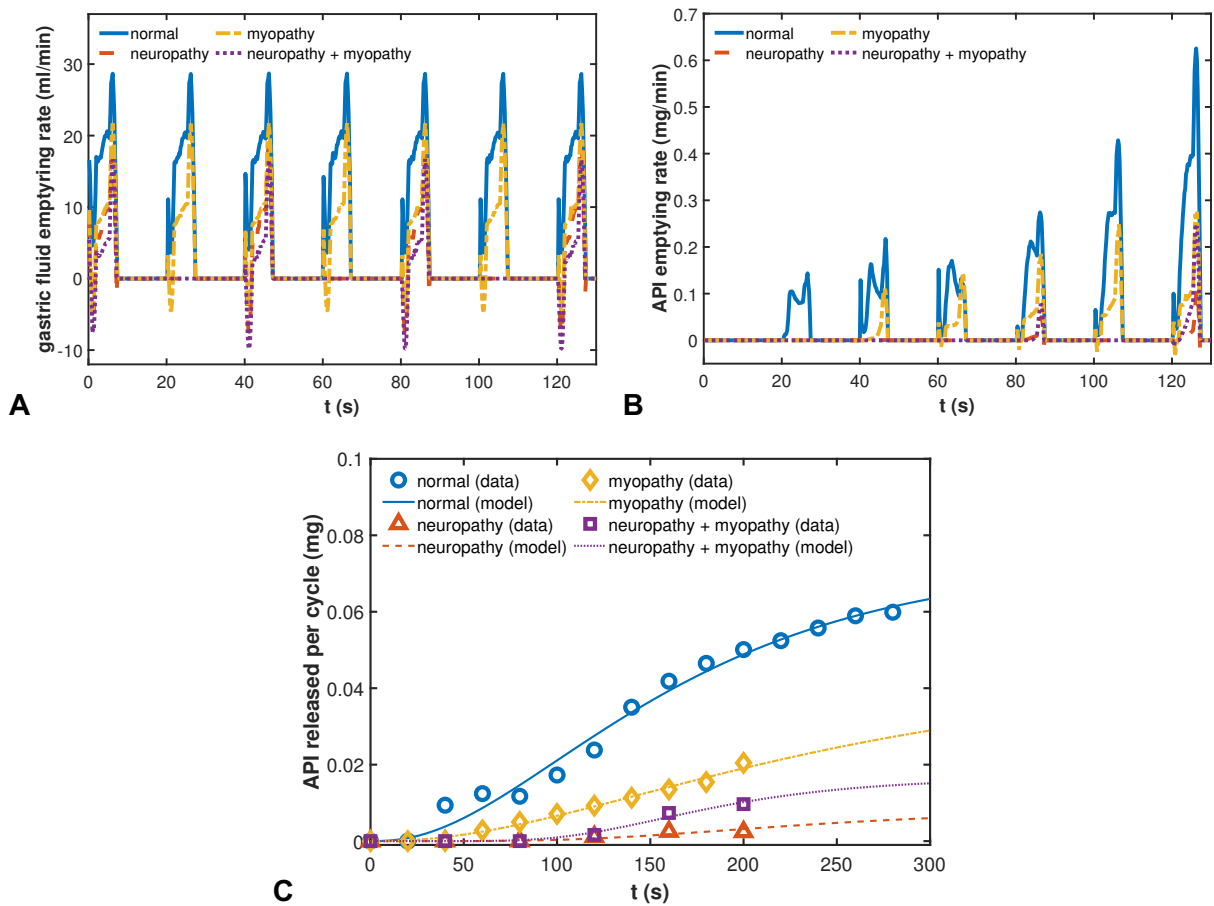

**Fig. 8.** Comparison for different types of gastroparesis. **A.** Quantification of gastric emptying rate into the duodenum measured by calculating the velocity flux through the pylorus. **B.** Quantification of the dissolved mass into the duodenum measured by calculating the concentration flux through the pylorus. **C.** Comparison of fit for a modified Elashoff model against simulation data for the amount of API emptied into the duodenum every cycle for different types of gastroparesis.

Fig. 8 summarizes the effect of altered gastric motility on dissolution and release of the API. Fig. 8A indicates that reductions in either the amplitude of the antral contraction or the frequency of antral contractions yield considerable decrease in gastric emptying rate. Moreover, Fig. 8B shows that there is also a delay in release of the dissolved API

|  | normal | neuropathy | myopathy | neuropathy+ myopathy |
|---|---|---|---|---|
| average dissolved API released | 0.016 mg every 20 sec | 0.00092 mg every 40 sec | 0.0051 mg every 20 sec | 0.0022 mg every 40 sec |

**Table 5.** Average dissolved active pharmaceutical ingredient (API) released into the duodenum per cycle for different gastric conditions. To calculate the average dissolved API released, the cycle-average of the flux of dissolved API in Fig. 7B is calculated, which is then integrated over the duration of one cycle that corresponds to the condition.

into the duodenum, as well as a significant reduction in the total amount of API in the duodenum.

Table 5 summarizes our findings in terms of average dissolved API released into the duodenum per cycle. To calculate this, we first obtain the cycle-average of the flux of dissolved API shown in Fig. 7B and then integrate this over the duration of one cycle that corresponds to the condition. We note that for the normal condition, 0.016 mg of API is released into the duodenum per cycle (20 sec), 0.00092 mg every 40 sec for neuropathy, 0.0051 mg every 20 sec for myopathy, and 0.0022 mg every 40 sec for both neuropathy and myopathy.

Fig. 8C compares the simulation data for the amount of API emptied into the duodenum every cycle for different types of gastroparesis. The data is then used to fit the modified Elashoff model (Eq. 2). The fitted parameters $\mathrm{API}_\infty$, $\alpha$, and $\beta$ are listed in Table 6. The predictions from the model confirms the earlier results that reductions in the amplitude or the frequency of the antral contractions yield significant decrease in the emptying of the API in long-term ($\mathrm{API}_\infty$) by up to 89%. Different gastric motilities also result in either decrease in the emptying rate or increase in the initial delay in emptying.

## 4. Discussion

|  | normal | neuropathy | myopathy | neuropathy+ myopathy |
|---|---|---|---|---|
| $\mathrm{API_\infty}$ (mg) | 0.0735 | 0.0083 | 0.0431 | 0.0165 |
| $\alpha$ | 0.0096 | 0.0106 | 0.0064 | 0.0169 |
| $\beta$ | 2.6004 | 7.6065 | 2.4855 | 13.8995 |

**Table 6.** Parameters obtained by fitting the modified Elashoff model (Eq. (2)) against the simulation data for the amount of API emptied into the duodenum every cycle for different types of gastroparesis (Fig. 8C). $\mathrm{API_\infty}$ is the predicted amount of API released per cycle at steady-state, and $\alpha$ and $\beta$ represent emptying rate and initial delay in emptying, respectively.

In this study, we demonstrate a computational model of drug dissolution in the human stomach and investigate the effect of posture and gastroparesis on drug dissolution and the emptying rate of the API into the duodenum. To the best of our knowledge, this is the first model that couples gastric biomechanics with pill movement and drug dissolution, and quantifies the API passing through the pylorus into the duodenum. With our model, we are able to calculate and compare the emptying rate and the release of dissolved API into the duodenum for a variety of physiological situations.

Previous studies have shown that the right lateral position leads to faster gastric emptying and that the left lateral and supine positions generally lead to slower mixing and gastric emptying of food[17,27]. It has been also known that sitting, standing, and recumbent right postures that accelerate emptying of the food also accelerate the absorption of orally administered drugs and thus the plasma drug concentration[27]. This study provides further insights into how posture impacts bioavailability of oral drugs. Our results agree with previous findings from clinical studies[3,28] that showed that the release of gastric content into the stomach and the bioavailability of oral drugs is maximized when the direction of gravity aligns with the antrum and the pylorus. Our results suggest that the location of the pill with respect to the core of the retropulsive jet is also an important factor in the dissolution and release of the API. These insights have important implications for accounting for posture in clinical studies of drug dissolution and to consider posture as a way of modulating the release of API concentration into the duodenum. This is particularly relevant for narrow absorption window (NAW) drugs, which are absorbed mainly in the upper part of the GI tract and require that pills be retained in the stomach longer compared to other oral drugs[15]. Moreover, our results indicate that there is significant difference in both the emptying rate and the amount of dissolved API into the duodenum, which is a crucial factor in determining the bioavailability of an oral drug. Moreover, our results show that the mean dissolved API released into the duodenum (0.023 mg) is of the same magnitude as the root mean squared deviation (0.019 mg) for the various postures. Indeed, we observe that certain postures can potentially reduce API bioavailability to a degree similar to that caused by a severe gastroparesis. Our long-term prediction using the Elashoff model shows that the amount of API being emptied can eventually be similar for three of the cases, different postures can lead to different rate at which enough API reaches the duodenum (Fig. 6C). This has an important implication especially for the rate of drug absorption for bedridden patients or elderly adults[9,27].

Our simulation results recapitulate what has been observed clinically for patients with gastroparesis and provide additional support for the utility of this in-silico modeling approach. In fact, our models of gastroparetic stomachs capture duodenogastric reflux, which has been known clinically to be associated with decreased MI[10,22]. The results also provide insights regarding the concept of the motility index (MI) and its clinical implications. For example, the neuropathic and the myopathic cases have similar values for MI, with the myopathic case being slightly smaller. However, in our simulations the myopathic case showed more efficient emptying of the dissolved API into the duodenum, indicating that MI does not fully encapsulate the functional severity of gastroparesis. Using the Elashoff model, we predict that this trend may continue long-term, owing to a significant increase in the initial delay in emptying for the neuropathic case. Another interesting example in which we show that our results may provide more mechanistic insights is the comparison between the neuropathic case and both neuropathic + myopathic case. Although MI for the neuropathic case is twice as large, we observe that the emptying of API is less efficient in this case, which is counterintuitive. This is explained by looking at the interplay between the strength of the ACW and the strength of retropulsive jet, and it is evident in our simulations that this balance is necessary in optimal emptying of API. We also use our model can predict that unlike being in different postures that showed more differences in the rate of emptying, gastroparetic stomachs have long-term effects on the actual amount of API emptied into the duodenum (Fig. 8C). Fig. 8 also shows that improving just one of the conditions may not eliminate the duodenogastric reflux and sufficiently increase the release of API concentration into the duodenum. Other approaches to improving gastric emptying without modifying the antral motility such as pyloroplasty and pyloromyotomy widen the opening at the pylorus but also increase the possibility of duodenogastric reflux[18]. Using this in-silico platform to compare different options to treat gastroparesis would be a future research topic that may provide clinical insight to improve guidelines for treatment selections.

One of the limitations of our model is the absence of tonic contraction, although we expect that only the overall emptying rate will change, and that the general behavior will be similar to our results. Due to the computational expense, the current simulations are limited to modeling a short duration of the dissolution process (about 3 minutes) which is quite short given that drug dissolution might occur over many hours. Methods to speed up the simulations are currently under development and these should allow for simulations that extend over O(1 hour). Despite these and other limitations, we have demonstrated that computational models and simulations of gastric fluid mechanics can provide useful and unique insights into the complex physiological processes that underlie drug dissolution. Further extensions of this modeling framework include coupling with physiologically-based pharmacokinetics to predict the absorption of the API in the duodenum, and multiphase flow modeling in the stomach to model more complex gastric contents.

## Acknowledgments

We acknowledge research funding from the NSF (Award CBET 2019405) and the NIH (Award 5R21GM139073-02). We also acknowledge the Johns Hopkins Discovery Award. J.H.L. also acknowledges the NIH TL1 Postdoctoral Biodesign Training Fellowship from the Johns Hopkins Institute for Clinical and Translational Research (Award TL1TR003100). This work used the Extreme Science and Engineering Discovery Environment (XSEDE), which is supported by the NSF (Award TG-CTS100002) and Startup Allocation (Award TG-MDE200001).

## Conflict of Interest

No conflict of interest.

**Supplementary Materials**

*A.1 Gastric motility model*

The motility of the stomach is modeled as the radial motion of the stomach lumen with respect to the centerline along the stomach,

$$\boldsymbol{x}_{\text{w}} = \boldsymbol{x}_{\text{w},0} + \lambda_{\text{a}}\boldsymbol{r},$$

in which $\boldsymbol{x}_{\text{w}}$ is the position vector of the lumen wall, $\boldsymbol{x}_{\text{w},0}$ is the initial position vector of the lumen wall, $\boldsymbol{r}$ is the vector from the wall to the antrum centerline, $\lambda_{\text{a}}$ is the wall strain, $s$ is the distance along the centerline of the stomach. See Fig. A1 for the schematic diagram of the implementation of gastric motility. Here, the centerline is divided into different regions; $s < s_1$: no antral contraction wave (ACW), $s < s_2$: ACW amplitude grows, $s_2 < s_3$: ACW amplitude is constant, $s_3 < s_4$: ACW amplitude increases (terminal antral contraction (TAC)), $s_4 < s_5$: ACW amplitude diminishes + segmental contraction of the distal antrum, and $s_4 < s_6$: TAC + pyloric closure and opening. This overall motion of the stomach lumen can be further divided into two motions,

$$\lambda_{\text{a}} = \lambda_{\text{a,ACW}} + \lambda_{\text{a,pyl}},$$

in which $\lambda_{\text{a,ACW}}$ is the wall strain from the ACW and $\lambda_{\text{a,pyl}}$ is the wall strain from the pyloric opening and closing.

The antral contraction consists of three phases: peristaltic contraction, terminal antral contraction (TAC), and antral relaxation[3]. The peristaltic contraction is modeled as the motion of the stomach lumen, which is prescribed by the propagation of the antral contraction wave (ACW),

$$\lambda_{\text{a,ACW}}(t,s) = \lambda_{\text{a,max}} \sum_n \frac{1}{2} \left[ \cos\left( \frac{2\pi \left( s - V_{\text{p}} t - n T_{\text{p}} V_{\text{p}} \right)}{W_{\text{p}}} \right) + 1 \right] h(s),$$

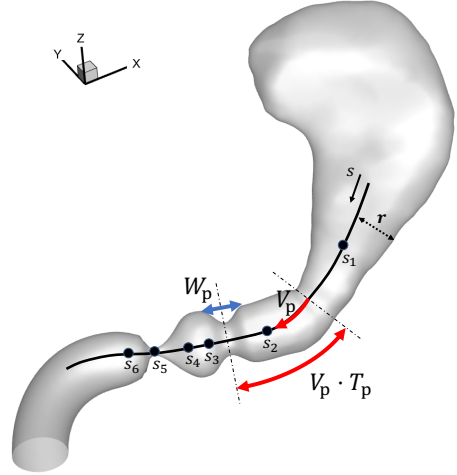

**Fig. A1.** Schematic diagram of the implementation of gastric motility. $s$ is the coordinate along the centerline, and the centerline is divided into different regions; $s < s_1$: no antral contraction wave (ACW), $s < s_2$: ACW amplitude grows, $s_2 < s_3$: ACW amplitude is constant, $s_3 < s_4$: ACW amplitude increases (terminal antral contraction (TAC)), $s_4 < s_5$: ACW amplitude diminishes + segmental contraction of the distal antrum, and $s_4 < s_6$: TAC + pyloric closure and opening. $V_{\text{p}}$ is the pulse propagation speed, $T_{\text{p}}$ is the pulse interval, $W_{\text{p}}$ is the pulse width.

$$h(s) = \begin{cases} 0, & s \leq s_1 \\ \frac{1}{2}\left(1 - \cos\left(\frac{s-s_1}{s_2-s_1}\pi\right)\right), & s_1 < s \leq s_2 \\ 1, & s_2 < s \leq s_3 \\ 1 + \frac{h_{\max}-1}{2}\left(1 - \cos\left(\frac{s-s_3}{s_4-s_3}\pi\right)\right), & s_3 < s \leq s_4 \\ \frac{h_{\max}}{2}\left(1 + \cos\left(\frac{s-s_4}{s_5-s_4}\pi\right)\right), & s_4 < s \leq s_5 \\ 0, & \text{otherwise} \end{cases},$$

in which $\lambda_{\text{a,max}}$ is the maximum diameter contraction ratio, $n$ is the pulse count, $V_{\text{p}}$ is the pulse propagation speed, $T_{\text{p}}$ is the pulse interval, $W_{\text{p}}$ is the pulse width, and $h(s)$ is a spatial damping function. For a healthy case, we use $V_{\text{p}} = 2.3$ mm/sec, $T_{\text{p}} = 20$ sec, $W_{\text{p}} = 20$ mm, and $\lambda_{\text{a,max}} = 0.45$ [1,2,5].

The opening and closing of the pyloric sphincter is prescribed by radial contraction and relaxation of the pyloric wall strain,

$$\lambda_{\text{a,pyl}} = p(s)\text{gate}(t),$$

in which $p(s)$ is defines the shape of the terminal antrum and the pylorus and $\text{gate}(t)$ deforms $p(s)$ to open and close the pylorus. Note that because here we also model the terminal antrum as well, so $\text{gate}(t)$ also deforms that region, which plays part in TAC. In our model, $p(s)$ and $\text{gate}(t)$ are defined as

$$p(s) = \begin{cases} \frac{1}{2}\left(1 - \cos\left(\frac{s-s_4}{s_5-s_4}\pi\right)\right), & s_4 \leq s \leq s_5 \\ \frac{1}{2}\left(1 - \cos\left(\frac{s-s_5}{s_6-s_5}\pi\right)\right), & s_5 < s \leq s_6 \\ 0, & \text{otherwise} \end{cases}$$

$$\text{gate}(t) = \begin{cases} A_c - \frac{A_c-A_o}{2}\left(1 - \cos\left(\frac{\tau}{T_{\text{opening}}}\pi\right)\right), & \text{pylorus opening} \\ A_o, & \text{pylorus open} \\ A_c - \frac{A_c-A_o}{2}\left(1 + \cos\left(\frac{\tau-T_{\text{open}}-T_{\text{opening}}}{T_{\text{opening}}}\pi\right)\right), & \text{pylorus closing} \\ A_c, & \text{pylorus closed} \end{cases},$$

in which $\tau = (t \bmod T_p)/T_p - T_0$ is the factional time in antral contraction cycle starting from when the pylorus begins to open $(T_0)$, $T_{\text{opening}} = 0.09$ is the fraction of time in one antral contraction during which the pylorus start to open (or close), $T_{\text{open}} = 0.17$ is the fraction of time during which the pylorus is kept open, and $A_c = 0.91$ and $A_o = 0.819$ are parameters that determine the closed and open configuration of the pylorus, respectively.

*A.2 Flow solver*

In this study, we use a sharp-interface immersed boundary solver ViCar3D[4] to simulate gastric fluid-structure interaction (FSI), which has been used and extensively validated for cardiovascular flows[6–9]. Here, we model the liquid content in the stomach as a Newtonian fluid, and we simulate the gastric flow by solving the incompressible Navier-Stokes equations,

$$\frac{\partial \boldsymbol{u}}{\partial t} + \boldsymbol{u} \cdot \nabla \boldsymbol{u} = -\frac{1}{\rho} \nabla p + \nu \nabla^2 \boldsymbol{u} + \boldsymbol{g},$$

$$\nabla \cdot \boldsymbol{u} = 0,$$

in which $\boldsymbol{u}$ is the flow velocity, $p$ is the static pressure of the fluid, $\rho$ is the fluid density, $\nu$ is the fluid kinematic viscosity, and $\boldsymbol{g} = -G\hat{\boldsymbol{z}}$ ( $G$ = 9.8 m/s$^2$ ) is the gravitational acceleration.

*2.3 Pill motion and dissolution*

The motion of the pill is obtained by solving 6-DOF equations of motion[10]

$$I \frac{\partial \boldsymbol{\omega}_{\mathrm{p}}}{\partial t} = \boldsymbol{M}_{\mathrm{f}} + \boldsymbol{M}_{\mathrm{c}},$$

$$m \frac{\partial \boldsymbol{u}_{\mathrm{p}}}{\partial t} = \boldsymbol{F}_{\mathrm{f}} + \boldsymbol{F}_{\mathrm{c}} + m\boldsymbol{g},$$

in which $I$ is the moment of inertia, $m$ is the mass, $\boldsymbol{u}_{\mathrm{p}}$ is the translational velocity, and $\boldsymbol{\omega}_{\mathrm{p}}$ is the angular velocity of the pill. $\boldsymbol{M}_{\mathrm{f}}$ and $\boldsymbol{F}_{\mathrm{f}}$ are the force and torque induced by shear stress and pressure of the surrounding fluid, $\boldsymbol{M}_{\mathrm{c}}$ and $\boldsymbol{F}_{\mathrm{c}}$ are the force and torque from contact with the stomach wall, and $\boldsymbol{g}$ is the gravitational acceleration.

The pill dissolution and release of API (denoted as $C_{\mathrm{API}}$) is treated as a scalar, and we compute it by directly solving the convection-diffusion equation

$$\frac{\partial C_{\text{API}}}{\partial t} + (\boldsymbol{u} \cdot \nabla) C_{\text{API}} = D \nabla^2 C_{\text{API}},$$

in which $D$ is the diffusion coefficient. Here, the local API concentration at the surface of the pill is equal to the solubility concentration of the pill ($C_{\text{API}} = C_{\text{s}}$).